# Monolayer-Defined Flat Colloidal PbSe Quantum Dots in Extreme Confinement

*Leon Biesterfeld,[a,b,c] Huu Thoai Ngo,[d] Ahmed Addad,[e] Wolfgang Leis,[f] Michael Seitz,[f] Gang Ji,[e] Bruno Grandidier,[d] Christophe Delerue,[d] Jannika Lauth,\*[a,b,c] Louis Biadala\*[d,e]*

a – Cluster of Excellence PhoenixD (Photonics, Optics, and Engineering – Innovation Across Disciplines), Welfengarten 1A, D-30167 Hannover, Germany.

b – Institute for Physical and Theoretical Chemistry, Eberhard Karls University of Tübingen, Auf der Morgenstelle 18, D-72076 Tübingen, Germany.

c – Institute for Physical Chemistry and Electrochemistry, Leibniz University Hannover, Callinstr. 3A, D-30167 Hannover, Germany.

d – Université de Lille, CNRS, Centrale Lille, Université Polytechnique Hauts-de-France, Junia-ISEN, UMR 8520-IEMN, F-59000 Lille, France.

e – Université Lille, CNRS, INRAE, Centrale Lille, UMR 8207-UMET-Unité Matériaux et Transformations, F-59000 Lille, France.

f – Institute for Inorganic Chemistry, Eberhard Karls University of Tübingen, Auf der Morgenstelle 18, D-72076 Tübingen, Germany.

ABSTRACT

Colloidal two-dimensional lead chalcogenide nanocrystals represent an intriguing new class of materials that push the boundaries of quantum confinement by combining a crystal thickness down to the monolayer with confinement in the lateral dimension. In particular flat PbSe quantum dots exhibit efficient telecommunication band-friendly photoluminescence (1.43 - 0.83 eV with up to 61% quantum yield) that is highly interesting for fiber-optics information processing. By using cryogenic scanning tunneling microscopy and spectroscopy, we probe distinct single layer-defined PbSe quantum dot populations down to a monolayer with in-gap state free quantum dot-like density of states, in agreement with theoretical tight binding calculations. Cryogenic ensemble photoluminescence spectra reveal mono-, bi-, and trilayer contribution, confirming the structural, electronic and theoretical results. From larger timescale shifts and ratio changes in the optical spectra we infer Ostwald ripening in solution and fusing in deposited samples of thinner flat PbSe quantum dots, which can be slowed down by surface passivation with $PbI_2$. By uncovering the interplay between thickness, lateral size and density of states, as well as the synthetic conditions and post-synthetic handling, our findings enable the target-oriented synthesis of two-dimensional PbSe quantum dots with precisely tailored optical properties at telecom wavelengths.

The sparse density of states (DOS) of (colloidal) quantum dots (QDs) is a direct function of the spatial charge carrier confinement in the materials and tunable *via* their size and shape.[1–4] Synthetic control over the dimensions of colloidal QDs allows for tailoring of their (photo-)physics and optoelectronic properties.[5,6] By controlling the anisotropic growth of QDs, e.g. in a one- or two-dimensional (1D, 2D) manner, an additional degree of freedom for tuning the band-edge energy and exciton binding energy (compared to zero-dimensional (0D) QDs) can be introduced.[7–9] For example, in colloidal 2D CdSe nanoplatelets (NPLs) with atomically precise thickness, the strong quantum confinement in the vertical dimension dominates the optical properties of the material, resulting in rapidly decaying (< 10 ns at room temperature (RT)) photoluminescence (PL) with narrow line widths below 40 meV.[10–12] Meanwhile, the lateral confinement in 2D CdSe enables few-nm-fine-tuning of the PL position (e.g. 459 - 463 nm for laterally confined 3.5 monolayer (ML) NPLs)[13] by reducing their length or width down to and below the exciton Bohr radius ($a_{B,\text{bulk CdSe}}$ = 5.6 nm,[14] $a_{B,3-5\text{ ML CdSe NPLs}}$ = 1 - 1.5 nm,[15,16] weakly depending on lateral size and thickness), thus obtaining CdSe NPLs with large aspect ratios (> 5.5)[13] or very small lateral extent (~1 nm),[17] exhibiting confinement in two or three dimensions.

Tuning the optical properties of 2D IV-VI semiconductors is equally interesting, since the materials exhibit tunable absorption and PL well within the telecommunication range[18–20] and are intriguing as classical or quantum emitters for applications in fiber optics and photonics.[21,22] Recently, colloidal synthesis protocols for cubic rock salt crystal structured 2D PbS, PbSe and PbTe have been established.[23–26] For example, in near-infrared (NIR) emitting colloidal 2D PbSe NPLs[24,25] (flat 2D PbSe QDs as will be shown here), the increased vertical confinement compared to spherical 0D PbSe QDs[18–20] results in excitonic absorption and PL at higher energies than in the 0D PbSe QDs (e.g. absorption at 1.74 eV for 2D PbSe QDs with a lateral size of 3 x 3 nm² compared to 1.24 eV for spherical PbSe QDs of 3 nm diameter).[25,27] Flat PbSe QDs are synthesized using lead oleate and selenourea precursors at 0°C, typically exhibit lateral sizes of 3 x 3 nm² to 7 x 7 nm² and show efficient PL (up to 61 % quantum yield (QY)) in the NIR between 1.43 - 0.83 eV (860 - 1510 nm). Due to a large bulk exciton Bohr radius of PbSe ($a_{B,bulk\ PbSe}$ = 46 nm),[28] flat PbSe QDs are quantum confined in all three dimensions ($x, y, z/a_B < 1$), with particularly strong confinement in the thickness dimension ($z/a_B \ll 1$). However, thorough experimental studies of charge carrier confinement in flat PbSe QDs, which are important for advancing the material toward optoelectronic applications, are missing up to now due to the scarce synthetic access to 2D PbSe samples with considerable PLQY. While optical measurements of nanocrystals (NCs) in solution give a straightforward view on ensemble-averaged quantum confinement of the samples (though limited to allowed transitions), directly probing the local electronic DOS is more complex and requires (cryogenic) scanning tunneling microscopy (STM) and spectroscopy (STS) methods. STS allows for determining excitation spectra of individual NCs, practically without any selection rules.[29] For instance, Ji *et al.*[3] and our group[30] have determined electronic band gaps of wet-chemically synthesized 2D CdSe NPLs, which are naturally larger than optical band gap values by the exciton binding energy, with varying thicknesses *via* STS and find larger electronic band gaps for thinner NPLs. Additionally, we reported on the occurrence of distinct Van Hove singularities in the conduction band of 2D CdSe NPLs that are directly correlated to their length, demonstrating the unique insights to be gained by combining STS with tight-binding (TB) calculations. For colloidal lead chalcogenide NCs, on the other hand, research has been focused on spherical 0D PbSe QDs,[31,32] PbSe/PbS core-shell QDs,[33] "molecular" aggregates of QDs,[31] and their epitaxially-fused superlattices up to now.[34–36] For example, Liljeroth *et al.* measured the DOS of individual spherical PbSe NCs (*d* = 3 - 7 nm) by STS, identified the $S_e$, $S_h$, $P_e$, and $P_h$ energy levels by semiquantitative comparison to TB calculations, and assigned the energy level separations to the transitions.[32]

Here we provide an in-depth study of extreme quantum confinement in colloidal 2D PbSe QDs by combining STM and STS with TB calculations and cryogenic ensemble PL of the samples. We find flat QDs with a thickness down to 0.6 nm, corresponding to a ML of cubic PbSe. Our measurements reveal band gaps free of in-gap states in 2D PbSe, and deduced DOS and electronic band gap values (1.67 eV, 1.26 eV, 1.00 eV) are assigned to mono-, bi-, and trilayer cubic PbSe using theoretical TB-calculated DOS, which unambiguously demonstrate the strong quantum confinement in the $z$-direction. By calculating the DOS for flat PbSe QDs of different lateral sizes we further show that lateral confinement in $x$- and $y$-directions, while not as extreme as the vertical confinement, can be used for tuning the size of the band gap over hundreds of meV. In perfect agreement with STS, three PL contributions are observed in cryogenic ensemble spectra, with their respective ratios changing for aged samples, suggesting an Ostwald ripening mechanism of thinner (1 and 2 ML) populations of flat PbSe QDs toward thicker (2 and 3 ML) QDs with smaller band gap in solution as well as a fusing mechanism in deposited samples. These findings advance the understanding of colloidal flat PbSe QDs by disentangling the vertical and lateral quantum confinement in these ultrathin structures and thus enable the synthetic targeting of specific flat QD dimensions resulting in tailor-made optical properties at technologically relevant wavelengths.

NIR emitting colloidal flat PbSe QDs stabilized by oleate and octylamine ligands were synthesized by a method described previously.[24,25] Figure 1a shows an overview HR-HAADF-STEM image of 2D PbSe QDs exhibiting shapes with slightly anisotropic lateral dimensions of (4.9 ± 1.1) x (3.9 ± 0.6) nm$^2$ and a corresponding small aspect ratio of 1.25 : 1 (see Figure S1a-c for typical images used for determining lengths and widths and a lateral size histogram from STEM). At higher magnification (Figure 1b) the lattice fringes of flat QDs are clearly visible. The measured lattice spacing of 3.1 Å (200) indicates the PbSe rock salt structure, supported by the corresponding FFT pattern (Figure 1c), which exhibits a set of diffraction peaks characteristic of cubic PbSe (lattice constant $a$ = 6.128 Å, space group $Fm\bar{3}m$, PDF card 01-077-0245). Figure 1d depicts a topographic large-scale STM image of the same flat QD sample shown in Figure 1a-c, drop-casted onto a gold substrate and annealed under high vacuum (see Figure 1e for a small-scale image of two flat QDs). Notably, the size and morphology of the flat QDs does not significantly change during annealing (see Figure S1 for a comparison of lateral size histograms from STEM (c) and STM (f)) and the QDs are highly stable under STM imaging conditions. For isolated flat PbSe QDs the height/QD thickness $z$ can be determined from STM images collected in constant current mode. We find multiple thicknesses ranging from 1.8 nm (Figure S2) down to 0.6 nm (Figure 1f), corresponding to a ML of cubic PbSe. We denote the ML thickness $z_{monolayer}$ as the height of the unit cell as the smallest/thinnest building block of PbSe, so that $z_{monolayer} \approx a_{cubic\ PbSe} \approx 0.6$ nm. Although PbSe is a non-layered isotropic material, ML PbSe has previously been theoretically considered as a topological crystalline insulator[37] and a suitable material platform for defect engineering.[38] Notably, Ekuma reported on the effect of vacancy defects on the electronic and optical properties of ML PbSe and find that defect-induced states predominantly reside outside the bandgap, rendering ML PbSe a rather defect-tolerant material.[38] However, up to now, experimental realizations of ML PbSe have been limited to epitaxial growth on van der Waals substrates (e.g. on $MoS_2$[39] and $VSe_2$[40]), which is challenging due to the induced strain between the crystal lattices of the two materials involved. To validate the STM height measurements and gain insight into the electronic properties of colloidal mono- to few-layer flat PbSe QDs, we discuss STS measurements collected on single flat QDs at 77 K in Figure 2.

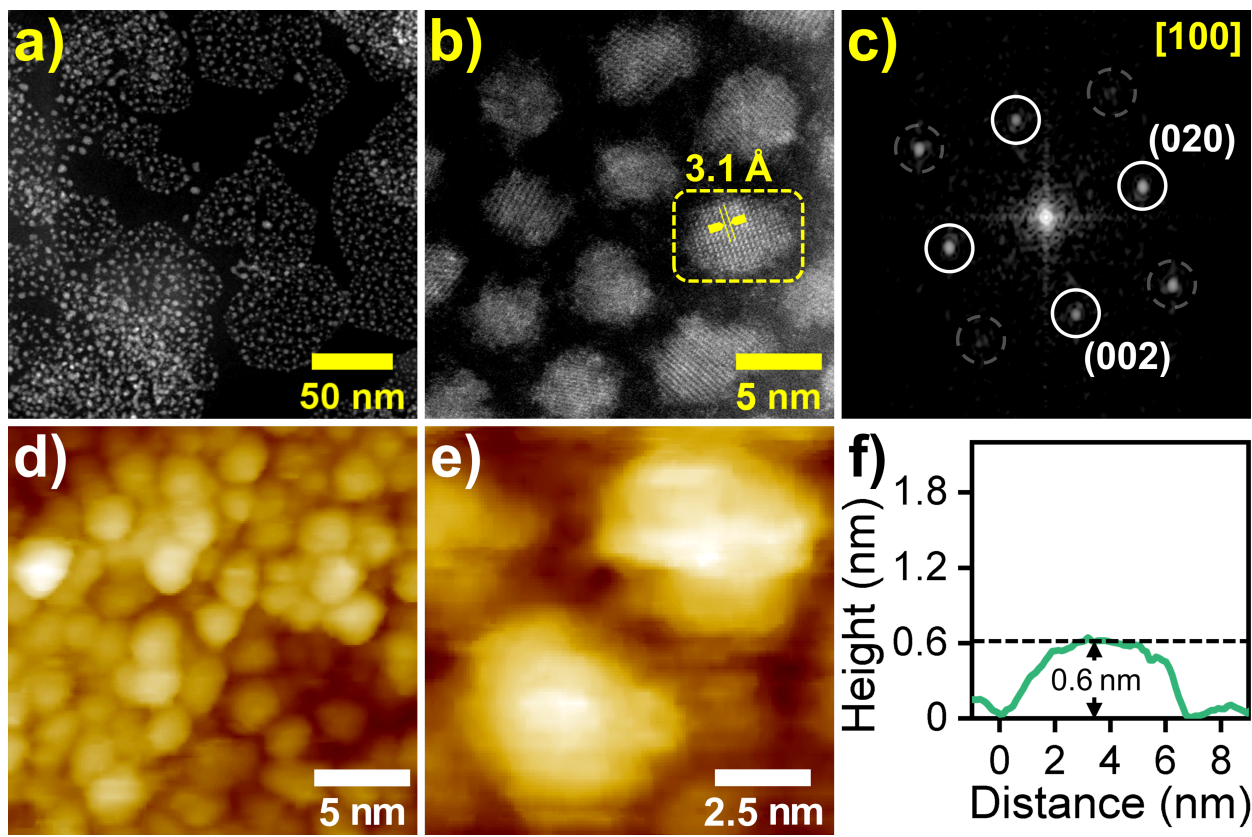

**Figure 1.** (a) Overview HR-HAADF-STEM image of flat PbSe QDs with average lateral dimensions of $(4.9 \pm 1.1) \times (3.9 \pm 0.6)$ nm$^2$. (b) HR image of individual crystalline 2D PbSe QDs (lattice spacing 3.1 Å (200)). (c) FFT pattern of the single flat QD region in (b), underpinning the crystallinity of the QD and exhibiting the characteristic diffraction peaks of cubic rock salt structured PbSe. (d) Low-magnification STM image of the flat PbSe QDs shown in (a), exhibiting average lateral sizes of 4.6 x 3.2 nm$^2$ after annealing under high vacuum. (e) Small-scale STM image of two flat PbSe QDs. (f) STM height profile of a flat QD with a thickness of 0.6 nm, corresponding to a monolayer of PbSe.

STS-measured $dI/dV$ spectra depend on the parameters of the double barrier tunnel junction formed between the STS tip, the sample, and the conductive substrate. In order to extract valid energy level structures and band gaps, resonant tunneling is required (shell-tunneling regime).[29,41,42] In this case, the tip-to-flat-2D-PbSe-QD junction limits the current, thereby preventing considerable charging of the sample. Figure 2a shows current dependent $dI/dV$ spectra collected on a thin flat PbSe QD; varying the set-point current $I_{set}$ changes the tip to QD distance and thereby the tunneling rate. Upon increasing $I_{set}$ from 100 pA to 500 pA, the peak positions do not shift, and the overall shape of the spectra remains constant, which implies resonant tunneling, and that the peaks align with the actual energy levels of the flat PbSe QDs, from which meaningful band gap values can be derived. In agreement with Overgaag et al.[31] we attribute the tolerance of PbSe to high set-point currents to the low effective charge carrier masses in PbSe, which result in spatially drawn-out orbitals, and a large tunneling coupling. For the flat PbSe QD shown in the STM image inset in Figure 2c, we find a zero-conductance region free of in-gap states, framed by single pronounced peaks on either side. The presence of sharp resonances in the $dI/dV$ spectra rather than a step-like energy level structure that would be expected for truly 2D quantum materials, provides evidence for the quantum confinement of 2D PbSe in all three dimensions ($x, y, z/a_B < 1$) and justifies the use of the term "flat QD". In accordance with literature, we assign the two peaks at negative and positive sample voltages to the $S_h$ and $S_e$ energy levels, resp.[32,43] The extracted electronic band gap of 1.65 eV is approx. 0.45 eV wider than the largest STS band gap reported for spherical PbSe QDs ($E_{g,STS} \approx 1.2$ eV for $d \approx 3.5$ nm) by Liljeroth et al., highlighting the extreme additional confinement in $z$ direction in flat PbSe QDs. Figure 2c depicts $dI/dV$ spectra obtained from four randomly selected spots within the same single flat PbSe QD (shown in Figure 2b). Although the expression of the valence (VB) and conduction band (CB) energy levels varies slightly between the different spots, we do not find any in-gap states at any of the probed spots. Furthermore, the band gap is consistent throughout the QD, underpinning that the deduced $E_{g,STS}$ values are valid.

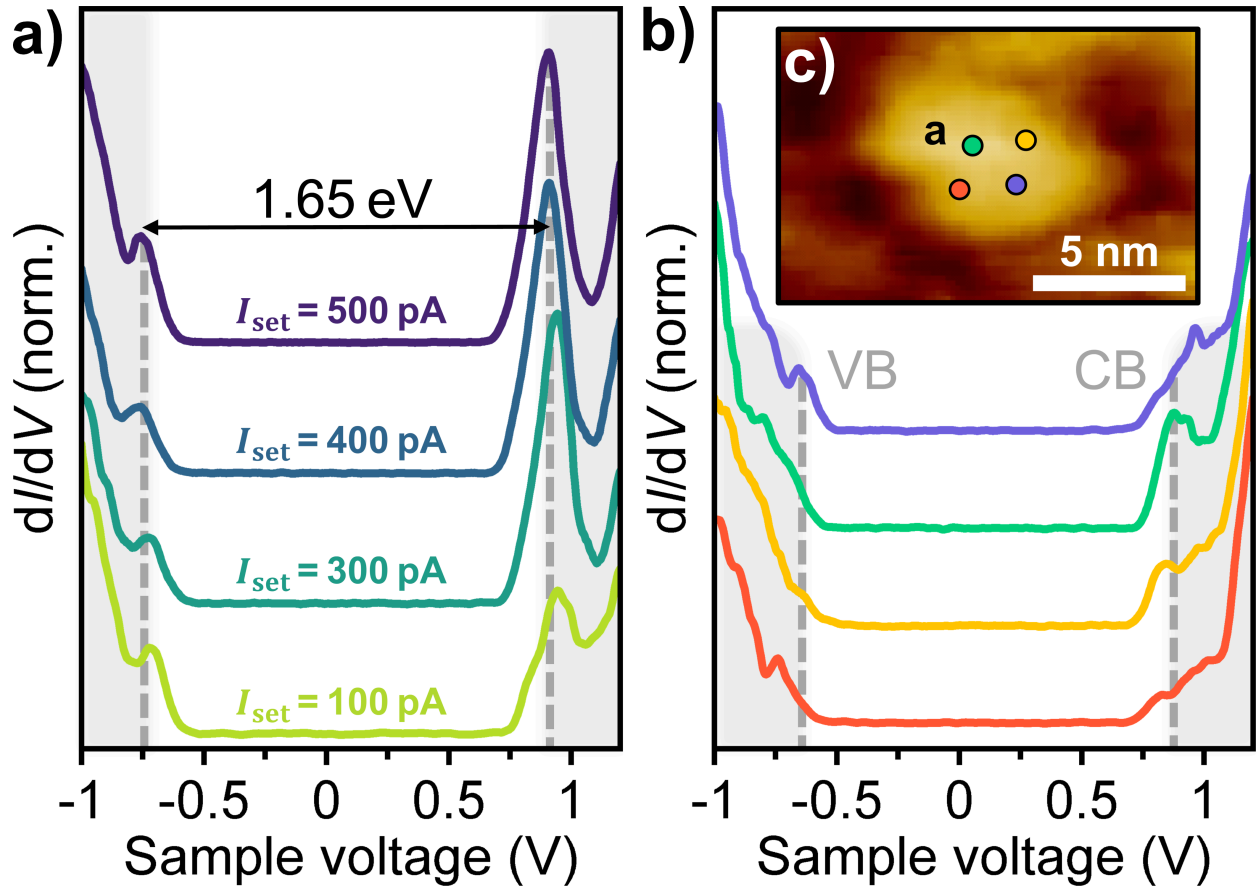

**Figure 2.** (a) Set-point current dependent STS spectra of an in-gap state-free flat PbSe QDs ($E_{g,STS}$ = 1.65 eV) showing sharp peaks that do not shift as $I_{set}$ is changed, implying resonant tunneling conditions. The occurrence of discrete peaks, instead of a step-like profile, reflects the quantum confinement of 2D PbSe QDs in all three dimensions. (b) d$I$/d$V$ spectra measured at four different spots within the same 2D QD (indicated in the STM image shown in (c)), which consistently show band gaps free of in-gap states with only minor fluctuations in the form of VB

and CB. (c) Small scale STM image of the flat PbSe QD studied by STS, spectra shown in (a) were collected at indicated location.

To further understand the correlation between the thickness/layer number of the flat QDs and their band gap size (as a measure for the degree of quantum confinement), we performed TB calculations to determine the theoretical DOS of mono- (0.6 nm) to trilayer (1.8 nm) flat PbSe QDs (with a lateral size of 5 x 5 nm$^2$, Figure 3a).[44] Figure 3b shows experimental data of three flat PbSe QDs with band gaps $E_{g,STS}$ of 1.00 eV, 1.26 eV, and 1.67 eV. The theoretical DOS quantitively replicates the experiment and we assign the measured DOS to mono-, bi-, and trilayer thick PbSe QDs, confirming the STM height measurement shown in Figure 1f spectroscopically. In addition, the good agreement with theoretical calculations confirms the crystalline quality of the synthesized flat QDs, as any vacancy defects or major shape irregularities would alter the measured DOS.

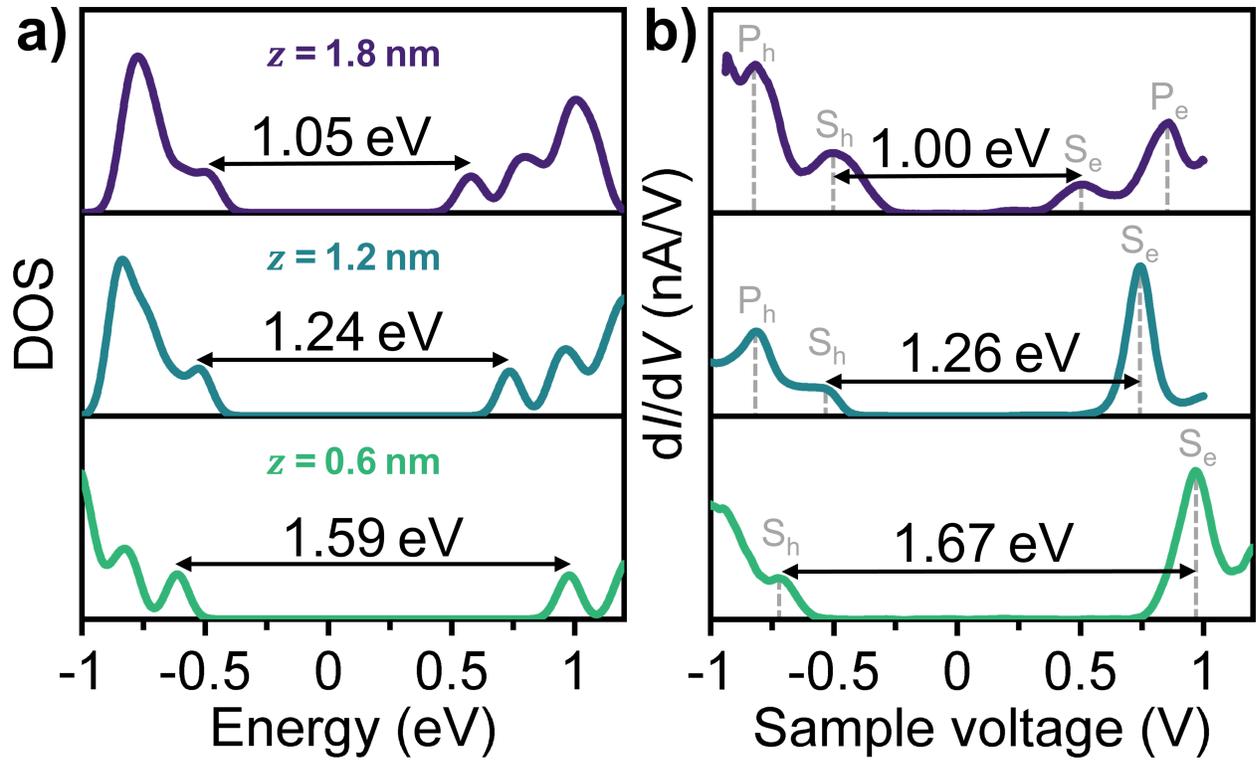

**Figure 3.** (a) Theoretical TB-calculated conduction and valence DOS of flat PbSe QDs with three different thicknesses. (b) Experimental STS spectra of in-gap state-free 2D PbSe QDs with three distinct band gaps collected at 77 K (tunneling conditions: $U_{sample}$ = 1.2 V, $I_t$ = 200 pA). By comparing theory to the experiment, we assign the measured DOS to monolayer ($E_g$ = 1.67 eV, $z$ = 0.6 nm), bilayer ($E_g$ = 1.26 eV, $z$ = 1.2 nm), and trilayer ($E_g$ = 1.00 eV, $z$ = 1.8 nm) cubic 2D flat PbSe QDs.

Figure 4a shows the room temperature ensemble PL and absorbance spectrum of flat PbSe QDs passivated with $PbI_2$ directly after synthesis.[25] Efficient PL is found centered around 1.34 eV (925 nm, PLQY of 43 %, increased from 23 % by $PbI_2$ treatment)[25] and excitonic absorption at 1.59 eV (780 nm). We note, the flat QDs depicted in Figure 4 are not from the same batch that was studied in STM/S (for completeness all optical data of the STM sample are shown in Figure S3).

We choose laterally smaller flat PbSe QDs with optical features at higher energies for the following cryo-PL discussion. Figure 4b shows PL of the same flat QD sample as in (a) at ~80 K, one year after synthesis (see Figure S4 for the corresponding room temperature PL spectrum at the time) We generally observe a bathochromic shift (by 0.24 eV) of aged samples, caused by aging of the QDs. The cryogenic PL spectrum centered around 1.10 eV (1127 nm) is best fitted by the sum of three Gaussians (accounting for inhomogeneous broadening by lateral size polydispersity). The higher and lower energy components centered around 0.99 eV and 1.19 eV (with associated FWHM of 139 meV and 138 meV, resp.) contribute equally (31 % and 30 %, resp.) to the total PL signal. The main contribution at 1.10 eV is significantly narrower with a FWHM of 76 meV, indicating a more uniform lateral size distribution. We attribute the three different PL contributions (0.99 eV, 1.10 eV, and 1.19 eV) to the three distinct flat QD thicknesses (1.8 nm, 1.2 nm, and 0.6 nm) observed by STM and STS. The width of the central narrow contribution is comparable to the FWHM typically obtained for spherical PbSe QDs (50 – 100 meV)[8,18] (albeit it should be noted that the FWHM of 2D PbSe[24,25] strongly depends on the spectral position) and all contributions are significantly narrower than typical values for 2D PbS NPLs at RT (> 200 meV)[21,23], indicating the potential of flat PbSe QDs for applications requiring narrow emission. Notwithstanding quantitative comparison to STS data due to it being a different sample, the difference between optical and electronic STS band gaps, points toward large exciton binding energies (hundreds of meV) in ultrathin PbSe QDs. Yang *et al.* calculated exciton binding energies $E_b$ of PbS and PbSe nanosheets (with infinite lateral dimensions) *via* an effective mass model (e.g. 80 meV for 3 nm thick PbSe),[45] that were experimentally reproduced by our group by THz spectroscopy on 2D PbS;[46] as a monolayer flat PbSe QD is only a fifth as thick and laterally confined we expect very tightly bound excitons with $E_b \gg 80$ meV. Notably, the ratio between the three PL contributions changes with longer aging times (see Figure S4 for a cryo-PL spectrum measured two months after the spectrum shown in Figure 4b), after another two months, we observe an increased population of the lower energy contributions, while the highest energy contribution decreases. This trend is also consistent for the PL of the flat QDs studied in STM/S (see Figure S3), hence we suggest a fusing mechanism of thinner (monolayer and bilayer) flat QDs toward thicker trilayer PbSe. We observe more pronounced aging in pristine flat PbSe QDs (see Figure S3 and S4) with a surface which is passivated by octylamine and oleate only, while $PbI_2$ passivation appears to slow down/hinder the fusing of flat QDs. These observations are consistent with previous reports for 2D CdSe NPLs by Dufour *et al.*,[47] who reported an increased colloidal stability for mixed halide- and octylamine-passivated 2D CdSe NPLs compared to purely organically passivated NPLs with a tendency to stack over time. Directly related to PbSe, Koh *et al.* found that oriented attachment of PbSe QDs is unfavored when $PbI_2/I^-$ is introduced, since the mismatch between Pb-I and Pb-Se bond lengths prevents the formation of bridges between adjacent NCs,[48] which likewise could hinder the vertical fusing of flat QDs in our case. Similarly, we have previously reported the necessity of $PbI_2$ passivation for the colloidal stabilization of 2D PbTe NPLs synthesized under comparable conditions.[26]

Figure 4c shows the TB-calculated correlation between the lateral size (assuming equal length and width, e.g. 3 x 3 $nm^2$) and the band gap of 2D PbSe QDs for mono- to trilayer thickness. The influence of lateral confinement starts to decrease from 4 nm length and width upwards, while the vertical confinement dominates the size of the band gap. This substantiates previous reports by us, in which the PL tunability of ultrathin PbSe[25] and PbTe[26] NPLs was assumed to be caused by a combination of both, thickness and lateral size, with thickness as the major factor. In addition, Figure 4c highlights the possibility and provides a target range to access the PL in the low attenuation telecommunication O-, E-, and S-bands, were glass fibers exhibit negligible losses, by further tailoring the dimensions of flat PbSe QDs.

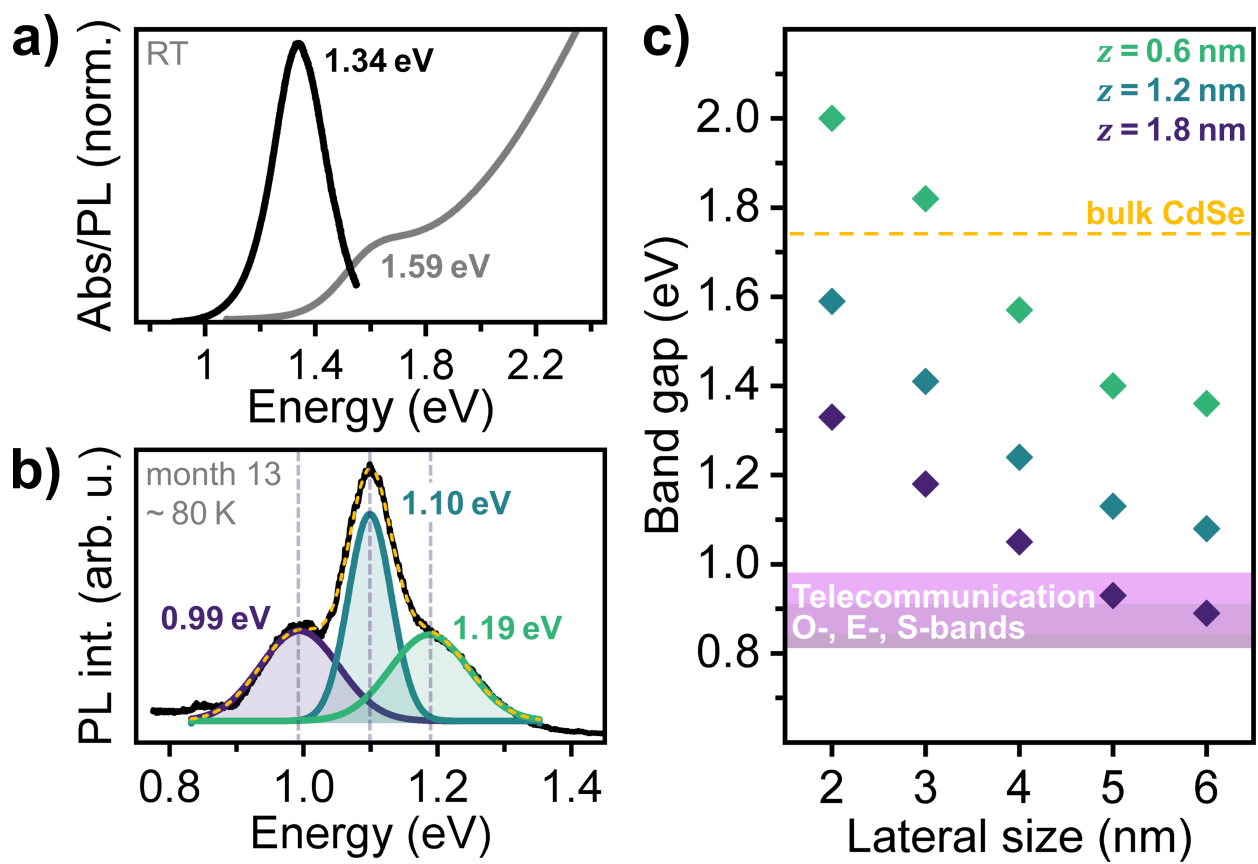

**Figure 4.** (a) Ensemble room temperature absorbance (grey) and PL spectrum (black) of colloidal PbI$_2$ passivated flat PbSe QDs, exhibiting excitonic absorption at 1.59 eV (780 nm) and NIR emission centered around 1.34 eV (925 nm) with a FWHM of 232 meV and a PLQY of 43 %. (b) Low-temperature PL spectrum of aged PbSe NPLs (shown in (a)) dip coated onto a magnesia stick and submerged in liquid N$_2$ (~80 K), revealing three distinct PL contributions, assigned to laterally polydisperse mono-, bi-, and trilayer flat PbSe QDs. (c) TB-calculated correlation between band gap and lateral size of mono- to trilayer 2D PbSe, highlighting the interplay of strong vertical

confinement and the minor lateral confinement in pushing 2D PbSe QD emission into the low attenuation telecommunication windows.

In conclusion, we have colloidally synthesized NIR-emitting 2D flat PbSe QDs with thicknesses down to a monolayer. We directly demonstrate the 2D nature of the QDs by STM (height) measurements and confirm the microscopy data by a comprehensive study of the electronic properties *via* low-temperature STS and TB calculations. In-gap state-free electronic band gaps of 1.67 eV, 1.26 eV, and 1.00 eV for mono-, bi-, and trilayers and a QD-like DOS with defined peaks on either side of the zero-conductance region, in agreement with theoretical DOS are demonstrated. Associated ensemble cryo-PL spectra are composed of three distinct contributions, attributed to the different thickness populations in 2D PbSe QDs and show consistent findings between structural, electronic and optical properties. From a shift of the PL contributions on longer times after synthesis, we infer a possible fusing mechanism of thinner 2D QDs to thicker populations with PL further in the NIR, which is prevented by $PbI_2$ surface passivation. Our study showcases the extreme quantum confinement in NIR-emissive ultrathin PbSe QDs and emphasizes their potential as innovative classical or quantum light sources for fiber optics-based photonics.

ASSOCIATED CONTENT

The following files are available free of charge.

Supporting Information:

ACKNOWLEDGMENT

Le. B. and J. L. gratefully acknowledge funding by the Deutsche Forschungsgemeinschaft (DFG, German Research Foundation) under Germany's Excellence Strategy within the Cluster of Excellence PhoenixD (EXC 2122, Project ID 390833453). J. L. is thankful for funding by the Ministry for Science and Culture of the State of Lower Saxony (MWK) for a Stay Inspired: European Excellence for Lower Saxony (Stay-3/22-7633/2022) Grant and for additional funding by an Athene Grant of the University of Tübingen (by the Federal Ministry of Education and Research (BMBF) and the Baden-Württemberg Ministry of Science as part of the Excellence Strategy of the German Federal and State Governments).

AUTHOR INFORMATION

**Corresponding Authors**

*jannika.lauth@uni-tuebingen.de and louis.biadala@iemn.fr

**Author Contributions**

Conceptualization: Le. B., Lo. B., and J. L.; investigation: Le. B., H. T. N., A. A., and C. D.; visualization and writing – original draft: Le. B.; Writing – review and editing: Le. B., B. G., Lo.

B., and J. L.; supervision: Lo. B. and J. L.; resources: M. S., C. D., Lo. B., J. L. All authors have read and given approval to the final version of the manuscript.

**Notes**

The authors declare no competing financial interest.

ABBREVIATIONS

0D, zero-dimensional; 1D, one-dimensional; 2D, two-dimensional; CB, conduction band; DOS, density of states; FFT, fast Fourier transform; FWHM, full width at half maximum; HR-HAADF-STEM, high-resolution high-angle annular dark-field scanning transmission microscopy; ML, monolayer; NC, nanocrystal; NIR, near-inrafred; NPL, nanoplatelet; PDF, powder diffraction file; PL, photoluminescence; QD, quantum dot; QY, quantum yield; RT, room temperature; STM, scanning tunneling microscopy; STS, scanning tunneling spectroscopy; TB, tight-binding; VB, valence band.